\documentclass[12pt]{article}

\usepackage{amsmath,amsmath,amssymb}
\usepackage{mathrsfs}
\usepackage{multirow}
\usepackage{graphicx}
\usepackage{authblk}
\usepackage{indentfirst}
\usepackage{multicol}  
\usepackage{tabu}
\usepackage{tabularx}
\usepackage{url}
\usepackage{fancyhdr}
\usepackage[numbers]{natbib}
\usepackage{lineno}
\usepackage{makecell}
\usepackage{fancyhdr}
\usepackage{colortbl}
\usepackage{caption}
\usepackage{float}
\usepackage[margin=1in]{geometry} 

\usepackage{setspace} 

\usepackage{hyperref}
\hypersetup{colorlinks=true,linkcolor=blue,anchorcolor=blue,citecolor=blue}

\title{
Topological Learning Prediction of Virus-like Particle Stoichiometry and Stability  
}

\author[1]{Xiang Liu} 
\author[2,3,4]{Xuefei Huang} 
\author[1,5,6]{Guo-Wei Wei \thanks{Corresponding author: weig@msu.edu}}
\affil[1]{Department of Mathematics, Michigan State University, MI, 48824, USA}
\affil[2]{Department of Chemistry, Michigan State University, MI, 48824, USA}
\affil[3]{Department of Biomedical Engineering, Michigan State University, MI, 48824, USA}
\affil[4]{Institute for Quantitative Health Science and Engineering, Michigan State University, MI, 48824, USA}
\affil[5]{Department of Electrical and Computer Engineering, Michigan State University, MI 48824, USA}
\affil[6]{Department of Biochemistry and Molecular Biology, Michigan State University, MI 48824, USA}
\date{}

\begin{document}
\maketitle

\paragraph{Abstract} 
Understanding the stoichiometry and associated stability of virus-like particles (VLPs) is crucial for optimizing their assembly efficiency and immunogenic properties, which are essential for advancing biotechnology, vaccine design, and drug delivery. However, current experimental methods for determining VLP stoichiometry are labor-intensive, and time consuming. Machine learning approaches have hardly been applied to the study of VLPs. To address this challenge, we introduce a novel persistent Laplacian-based machine learning (PLML) mode that leverages both harmonic and non-harmonic spectra to capture intricate topological and geometric features of VLP structures. This approach achieves superior performance on the VLP200 dataset compared to existing methods. To further assess robustness and generalizability, we collected a new dataset, VLP706, containing 706 VLP samples with expanded stoichiometry diversity. Our PLML model maintains strong predictive accuracy on VLP706. Additionally, through random sequence perturbative mutation analysis, we found that 60-mers and 180-mers exhibit greater stability than 240-mers and 420-mers. 
 
\paragraph{SIGNIFICANCE}
We proposed a topological Laplacian-based machine learning model for analyzing the stoichiometry and stability of virus-like particles (VLPs), which are critical factors in rational vaccine design and drug delivery. Our model demonstrated superior performance on the VLP200 dataset compared to competing methods. We have also curated new datasets for stoichiometry prediction. We introduced sequence mutational perturbation to analyze the stability of VLPs, revealing the higher stability of the 60-mer and 180-mer VLPs than 240-mers and 420-mers for the first time. The proposed approaches can be applied to viral capsid analysis and vaccine design.  
 
  
\newpage
	 
\section{Introduction}
Virus-like particles (VLPs) are highly immunogenic macromolecules that self-assemble from viral structural proteins \cite{chang2022structural}. They resemble or mimic the structure, versatility, size, and symmetry of the native viruses from which they are derived but lack viral nucleic acids, rendering them non-infectious and thus safer candidates for vaccine development \cite{chackerian2007virus,kushnir2012virus,roldao2010virus}. VLPs have been widely employed in the development of human vaccines against various viruses, and several VLP-based vaccines, such as those for hepatitis B \cite{mcaleer1984human} and human papillomavirus \cite{bryan2016prevention}, have been successfully developed, approved by the FDA, and commercialized. 
A critical determinant of VLP functionality is their stoichiometry, defined as the number of protein subunits comprising the particle. 
The stoichiometry of VLPs plays a pivotal role in determining their assembly efficiency and modulating their immunogenic properties  (CPs) \cite{pushko2013development,qin2023precise}.

Consequently, accurate prediction of VLP stoichiometry is essential for rational vaccine design and optimization. 
Several experimental methods, including analytical ultracentrifugation \cite{lebowitz2002modern}, light scattering methods \cite{some2013light}, and mass photometry \cite{young2018quantitative}, have been used to determine the VLP stoichiometry. However, these experimental methods require purified protein samples and are time-consuming, making them unsuitable for large-scale stoichiometry analysis. Consequently, there is a pressing need to develop computational approaches to predict VLP stoichiometry. 

Constructing reliable datasets is the first step in developing computational methods, especially machine learning models, for predicting VLP stoichiometry. However, there is only one public available dataset, termed VLP200 \cite{zhang2025classifying}, for this propose as far as we know. The VLP200 dataset comprises 200 protein sequences, equally divided into VLPs with 60-mer and 180-mer stoichiometry. Its construction is based on the authors’ observation that most homomeric VLPs tend to adopt either 60-mer or 180-mer configurations, as supported by a preliminary analysis of VLP entries from the RCSB Protein Data Bank (PDB) \cite{berman2000protein}. 
The dataset also aligns with the principles of the Caspar-Klug theory (CK theory) \cite{caspar1962physical}, which assumes that most viruses adopt icosahedral symmetry, and that their structural organization is determined by the triangulation number $T$, where a $T$-number corresponds to $60T$ protein subunits. Thus, the 60-mer and 180-mer VLPs in the dataset correspond to $T=1$ and $T=3$, respectively. This ensures VLP200 as a reliable benchmark for evaluating computational methods.
However, the VLP200 dataset remains limited in scope and requires further expansion to include a broader range of $T$-number classes ($T=4,7,9,13,16$) and a larger number of samples. Although the developers of the VLP200 dataset have developed several machine learning models for stoichiometry prediction, the performance remains limited, with the highest reported AUC reaching only 0.84 under 10-fold cross-validation. This highlights the need for innovative modeling strategies to improve predictive accuracy.

Persistent homology, a key theory in topological data analysis (TDA) \cite{carlsson2009topology}, has been integrated with machine learning for protein classification in computational biology \cite{cang2015topological}. Notably, TDA was first combined with deep neural networks, leading to topological deep learning (TDL) model in 2017 \cite{cang2017topologynet}. TDL is the new frontier in rational learning \cite{papamarkou2024position}. 
This approach demonstrated superior performance in predicting protein-ligand binding affinities and impacts of protein mutations \cite{cang2017topologynet}. Most impressively, persistent homology-based methods achieved victories in D3R Grand Challenges, a global competition series in computer-aided drug design \cite{nguyen2019mathematical,nguyen2020mathdl}. These successes highlight the effectiveness of persistent homology-based approaches for analyzing intrinsically complex multiscale data, such as biomolecular structures. 
However, persistent homology has intrinsic limitations, particularly in its ability to capture geometric variations that do not alter homotopy types. To address these limitations, persistent spectral graph, also known as persistent Laplacian, has been introduced by Wang et al. \cite{wang2020persistent}. Persistent Laplacian captures topological features through its harmonic spectra (i.e., persistent homology) and simultaneously encodes rich geometric characteristics via its non-harmonic spectra \cite{wang2020persistent}. Extensive evaluations on more than 30 protein engineering datasets have demonstrated that persistent Laplacians consistently outperform persistent homology \cite{qiu2023persistent}. Remarkably, persistent Laplacians have enabled the early prediction of emerging dominant SARS-CoV-2 variants, such as Omicron BA.4 and BA.5, two months prior to their announcement by the World Health Organization (WHO) \cite{chen2022persistent}. A survey of various persistent topological Laplacians is available \cite{wei2025persistent}. Recent advances in TDA and TDL are also reviewed \cite{su2025topological}. Despite their success in biomolecular data analysis, TDL or TDA-based methods, particularly those leveraging persistent Laplacians, have not been applied to VLP analysis.

In this work, we propose a persistent Laplacian-based topological machine learning (PLML) model for predicting the stoichiometry and stability of VLPs. In PLML, the VLP structure is represented as filtered simplicial complexes, including Vietoris-Rips complex and Alpha complex, which provides a multiscale characterization of the structural information within the data. From these topological representations, we compute persistent Laplacians to extract informative topological features. These topological features are then used as input for a gradient boosting tree algorithm to perform the prediction task. We evaluate PLML on the VLP200 dataset, the results demonstrate that PLML significantly outperforms other existing methods across all evaluation metrics. To further assess the model’s robustness and generalizability, we curate an expanded dataset, VLP706, comprising 706 VLP samples with four stoichiometries: 60-mer, 180-mer, 240-mer, and 420-mer. On this extended dataset, PLML achieves an accuracy of 0.858 and an AUC of 0.956 in 10-fold cross-validation, demonstrating its strong predictive performance. The proposed model is utilized to analyze the stability of VLPs by randomly perturbing their sequences with up to 50\% in the amino acid context.

\section{Results}
\subsection{Overview of the PLML Model}
\begin{figure}[ht]
    \centering
    \includegraphics[width=1\textwidth]{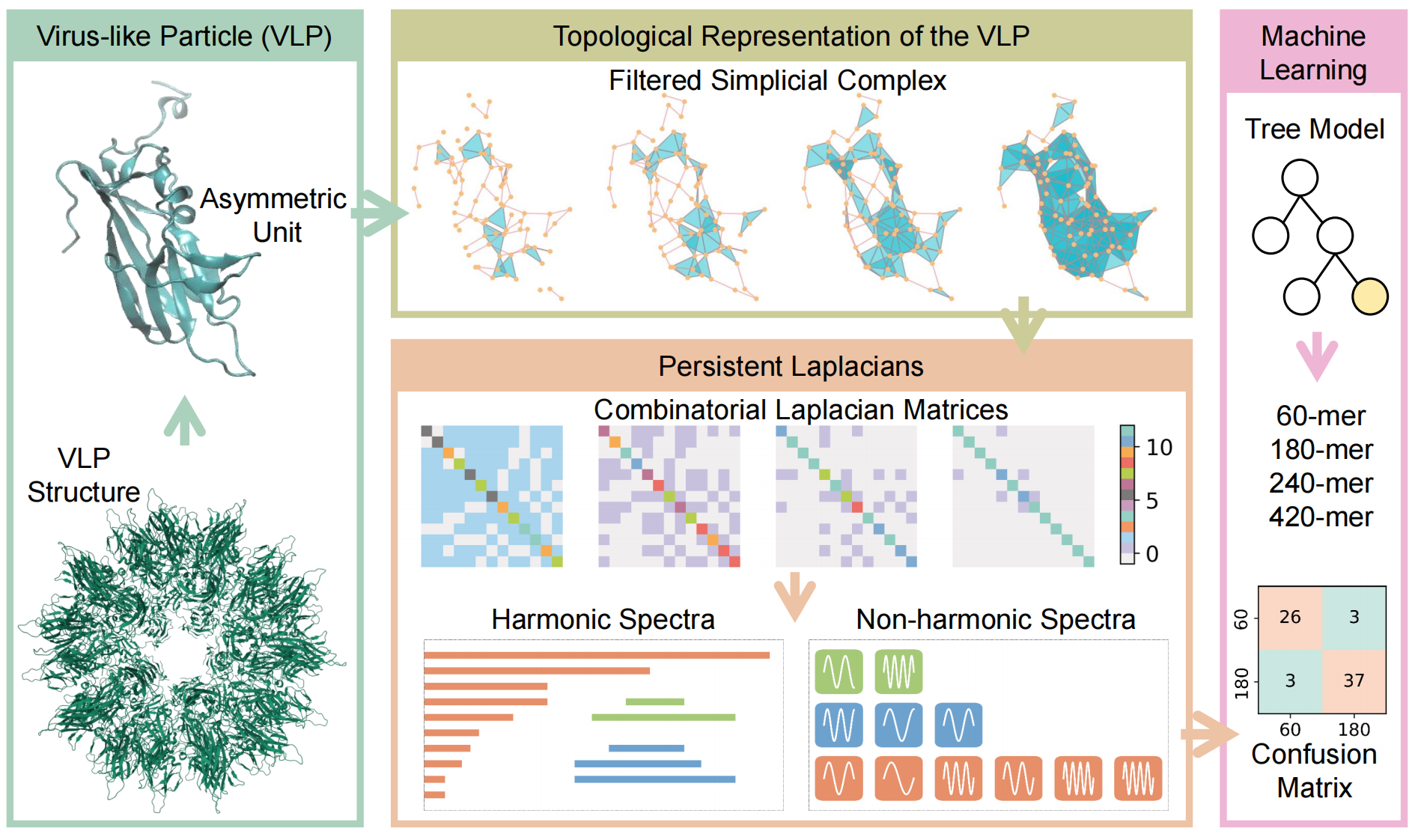}
    \caption{Illustration of the PLML model architecture. For each virus-like particle, its asymmetric unit is extracted as the region of interest to reduce complexity while preserving structural integrity. A filtered simplicial complex is then constructed from this unit to provide a topological representation of the data. Persistent Laplacians are subsequently computed from these complexes to extract multiscale topological features. Finally, these features are used as input to a gradient boosting tree algorithm for the stoichiometry classification task.   }
    \label{fig:model}
\end{figure}
The overall workflow of our model is illustrated in Figure~\ref{fig:model}. Given a virus-like particle, its asymmetric unit is first extracted as the region of interest to reduce computational complexity while preserving essential structural information. Next, filtered simplicial complexes, including the Vietoris-Rips complex and the Alpha complex, are constructed to provide a topological representation of the extracted units. These constructions form a sequence of nested simplicial complexes that capture structural patterns of the data at various scales. Based on these multiscale topological representations, persistent Laplacians are computed to extract topological features. The harmonic spectra of the persistent Laplacians correspond precisely to persistent homology that captures the topological structural information, while the non-harmonic spectra encode additional geometric and combinatorial properties of the data. Finally, the resulting topological features are used as input to a machine learning (gradient boosting tree \cite{friedman2001greedy}) algorithm for training and prediction in the classification of VLP stoichiometry.

\subsection{Evaluation of the Model}

\subsubsection{Datasets}
We consider a recently curated dataset comprising 200 VLP protein sequences, denoted by VLP200, with an equal number of sequences (100 each) corresponding to the stoichiometries of 60-mer and 180-mer. The authors of \cite{zhang2025classifying} collected this dataset through the advanced search function from the RCSB PDB, resulting in a binary classification task that distinguishes VLP sequences as either 60-mer or 180-mer. In our implementation, the structural subunits of the VLP200 dataset are constructed using AlphaFold2 and subsequently used for topological feature generation.

To enable a more robust and reliable evaluation of our model, we construct a bigger dataset consisting of 706 VLP protein structures, denoted by VLP706, corresponding to four common stoichiometries: 60-mer, 180-mer, 240-mer, and 420-mer. These structures are retrieved from the RCSB PDB using its advanced search function. Specifically, we set the ``Full Text'' field to ``virus-like particle'', select the number of protein instances (chains) per assembly as 60, 180, 240, and 420 for 60-mer, 180-mer, 240-mer, and 420-mer respectively under the ``Assembly Features'' section of ``Structure Attributes'', and specify the ``Symmetry Type'' as icosahedral.
This search yields 210 entries for 60-mer, 261 entries for 180-mer, 188 entries for 240-mer, and 47 entries for 420-mer VLPs. Other stoichiometries, such as 540-mer, 780-mer, and 960-mer, have fewer than 20 entries each and are therefore excluded due to their limited sample sizes. As a result, we construct a new multi-class dataset consisting of $706=210+261+188+47$ VLP samples. In this new dataset, the structural subunits are collected from the RCSB PDB, while AlphaFold2 is employed to model those samples lacking experimentally determined structures.

\subsubsection{Evaluation Protocols}
The Area Under the Receiver Operating Characteristic Curve (AUC), sensitivity, specificity, precision, negative predictive value (NPV), and accuracy (ACC) are employed as evaluation metrics to facilitate a fair comparison with benchmark methods reported in the literature \cite{zhang2025classifying}. To enhance the reliability of the results, we independently repeated the process 50 times with different random seeds and used the average value of all 500 runs (the strategy used in \cite{zhang2025classifying}) as the final performance of our model.

\subsubsection{Performance on the Existing Dataset}
\begin{figure}[ht]
    \centering
    \includegraphics[width=1\textwidth]{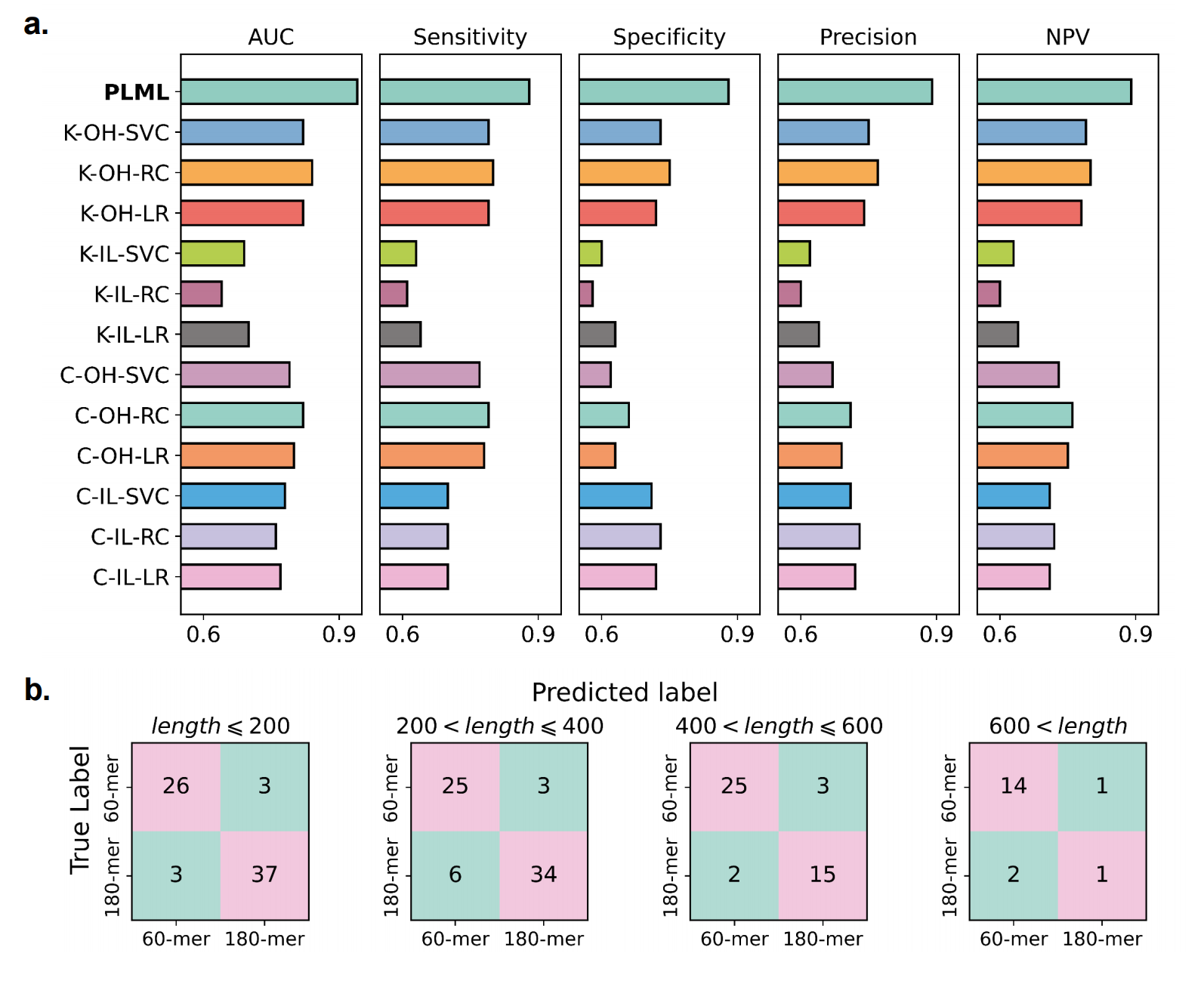}
    \caption{Illustration of model performance on the VLP200 dataset. (a): performance comparison between our model and other existing models in terms of AUC, sensitivity, specificity, precision, and NPV. The results demonstrate that our model consistently outperforms existing methods across all evaluation metrics. (b): Confusion matrix representations of our model’s performance across four groups of VLP sequences stratified by sequence length. These results highlight the model’s robustness across various sequence lengths. }
    \label{fig:VLP200-result}
\end{figure}
We evaluate our model on the VLP200 dataset using 10-fold cross-validation to ensure robust performance assessment and enable a fair comparison with existing methods. The results of our model and other competing approaches are presented in Figure~\ref{fig:VLP200-result}~(a). It can be seen that our model consistently outperforms all other methods across all five evaluation metrics, including AUC, sensitivity, specificity, precision, and NPV as defined in Section S3 of the Supplementary Information. Specifically:

\begin{itemize}
\item Our model achieves an AUC of 0.94, outperforming the second-best model, which attains 0.84.
\item Our model obtains a sensitivity of 0.88, compared to 0.80 for the second-best method.
\item Our model reaches a specificity of 0.88, notably higher than the 0.75 achieved by the next best approach.
\item Our model achieves a precision of 0.89, surpassing the 0.77 of the second-best model.
\item Our model yields an NPV of 0.89, whereas the second-best model records 0.80.
\end{itemize}

Notably, our model achieves scores of at least 0.88 across all five evaluation metrics. These consistently superior results highlight the effectiveness and robustness of our model in VLP stoichiometry classification. More details of our results are given in Table \ref{tab:VLP200-result}. 

\begin{table}[h]
\centering
\caption{Performance of PLML and existing methods on the VLP200 dataset. The results are reported as mean $\pm$ standard deviation over 50 runs. The best results are in bold. The benchmark results are derived from the literature \cite{zhang2025classifying}.   }
\label{tab:VLP200-result}
\begin{tabular}{l| c| c| c|c|c }
    \hline
    Method & AUC&Sensitivity&Specificity & Precision & NPV \\
    \hline
    C-IL-LR & 0.77$\pm$0.11 & 0.70$\pm$0.14&0.72$\pm$0.13  & 0.72$\pm$0.11   &0.71$\pm$0.12  \\
    \hline
    C-IL-RC &0.76$\pm$0.11 &0.70$\pm$0.16&0.73$\pm$0.14 & 0.73$\pm$0.12 &0.72$\pm$0.12 \\
    \hline
    C-IL-SVC & 0.78$\pm$0.12&0.70$\pm$0.15 &0.71$\pm$0.14& 0.71$\pm$0.11 &0.71$\pm$0.12\\
    \hline
    C-OH-LR &0.80$\pm$0.10 &0.78$\pm$0.12 &0.63$\pm$0.15&0.69$\pm$0.10&0.75$\pm$0.12\\
    \hline
    C-OH-RC &0.82$\pm$0.09 &0.79$\pm$0.12 &0.66$\pm$0.15&0.71$\pm$0.11&0.76$\pm$0.12\\
    \hline
    C-OH-SVC &0.79$\pm$0.11&0.77$\pm$0.13 &0.62$\pm$0.15&0.67$\pm$0.10&0.73$\pm$0.13  \\
    \hline
    K-IL-LR &0.70$\pm$0.11 &0.64$\pm$0.15 &0.63$\pm$0.15&0.64$\pm$0.11&0.64$\pm$0.10 \\
    \hline
    K-IL-RC &0.64$\pm$0.12 &0.61$\pm$0.16 &0.58$\pm$0.16&0.60$\pm$0.11&0.60$\pm$0.12 \\
    \hline
    K-IL-SVC &0.69$\pm$0.12&0.63$\pm$0.17 &0.60$\pm$0.17&0.62$\pm$0.13&0.63$\pm$0.13 \\
    \hline
    K-OH-LR &0.82$\pm$0.09 &0.79$\pm$0.12 &0.72$\pm$0.14&0.74$\pm$0.11&0.78$\pm$0.11 \\
    \hline
    K-OH-RC &0.84$\pm$0.08 &0.80$\pm$0.12 &0.75$\pm$0.14&0.77$\pm$0.10&0.80$\pm$0.11 \\
    \hline
    K-OH-SVC&0.82$\pm$0.09 &0.79$\pm$0.12 &0.73$\pm$0.14&0.75$\pm$0.10&0.79$\pm$0.11\\
    \hline
    \textbf{PLML} &\textbf{0.94$\pm$0.05}&\textbf{0.88$\pm$0.10} & \textbf{0.88$\pm$0.11} & \textbf{0.89$\pm$0.09} & \textbf{0.89$\pm$0.09} \\
    \hline
\end{tabular}
\end{table}

Furthermore, We analyze the model performance across different sequence lengths based
on the predictions from 10-fold cross-validation.
We divide the 200 VLP protein sequences into four groups based on their sequence length $n$: 69 sequences with $n\leq 200$, 68 with $200<n\leq 400$, 45 with $400<n\leq 600$, and 18 with $n>600$. The confusion matrix representation of the results is shown in Figure~\ref{fig:VLP200-result}~(b), our model achieves classification accuracies of 0.913, 0.868, 0.889, and 0.833 for these respective groups. The performance remains consistently high across the first three groups, with a slight decline observed in the group with $n>600$. This decrease is likely due to the small sample size in that group, which may limit the model’s ability to generalize effectively. These results demonstrate that our model maintains robust performance across VLPs with a broad range of sequence lengths.

\subsubsection{Performance on the New Dataset}
The initial VLP200 dataset comprises only 200 VLP protein sequences, which limits the comprehensiveness of model evaluation, despite our model achieving the best performance on this benchmark. To enable a more robust and extensive assessment, we constructed a larger dataset, VLP706, consisting of 706 VLP samples retrieved from the RCSB PDB. This expanded dataset includes four stoichiometry classes: 210 entries for 60-mers, 261 for 180-mers, 188 for 240-mers, and 47 for 420-mers. The increased sample size and label diversity offer a more reliable and challenging benchmark for evaluating computational models in VLP stoichiometry classification.

We first evaluate our model on the binary classification task involving 60-mer and 180-mer samples (210 and 261 entries, respectively). The model achieves an AUC of 0.98, sensitivity of 0.96, specificity of 0.95, precision of 0.96, and NPV of 0.95. For comparison, the corresponding performance on the VLP200 dataset was 0.94 (AUC), 0.88 (sensitivity), 0.88 (specificity), 0.89 (precision), and 0.89 (NPV). This significant performance gain suggests that our topological features can not only capture the intrinsic complex structural distinctions between 60-mer and 180-mer VLPs but also effectively utilize additional data to enhance generalization.


\begin{table}[h]
\centering
\caption{Performance of our PLML model on the VLP706 dataset, evaluated through binary, 3-class, and 4-class classification tasks.  }
\label{tab:VLP706-result}
\begin{tabular}{l| c| c| c|c|c }
    \hline
    Task & binary&binary&binary & 3-class & 4-class \\
    \hline
    Label & 60,180 & 60,240&180,240  & 60,180,240   &60,180,240,420  \\
    \hline
    AUC & 0.983&0.993&0.925 & 0.956 &0.956 \\
    \hline
    ACC & 0.954&0.962 &0.864& 0.877 &0.858\\
    \hline
\end{tabular}
\end{table}

We further extend the evaluation by incorporating the 240-mer and 420-mer classes, performing binary, 3-class, and 4-class classification tasks. The 420-mer samples are excluded from the pairwise binary classification due to their small sample size. The results, summarized in Table~\ref{tab:VLP706-result}, demonstrate that our model maintains high performance in terms of AUC across increasingly complex classification settings. Although the classification accuracy declines slightly as the number of classes increases, it remains strong, with a minimum of 0.858 in the 4-class task. Interestingly, the model's performance across the binary classification tasks follows a logical trend: classification between 60-mer and 240-mer achieves the highest performance, followed by 60-mer vs. 180-mer, and finally 180-mer vs. 240-mer. This ranking aligns with the structural similarities among the classes, as 180-mer and 240-mer share the most resemblance, whereas 60-mer and 240-mer exhibit the largest structural differences. Note that the 420-mer class is underrepresented, with only 47 samples, introducing a class imbalance in the dataset. Despite this challenge, our model continues to perform well, demonstrating its robustness and effectiveness for VLP stoichiometry analysis. 


\subsection{Structural Stability under Sequence Mutational Perturbation}
\begin{figure}[ht]
    \centering
    \includegraphics[width=0.9\textwidth]{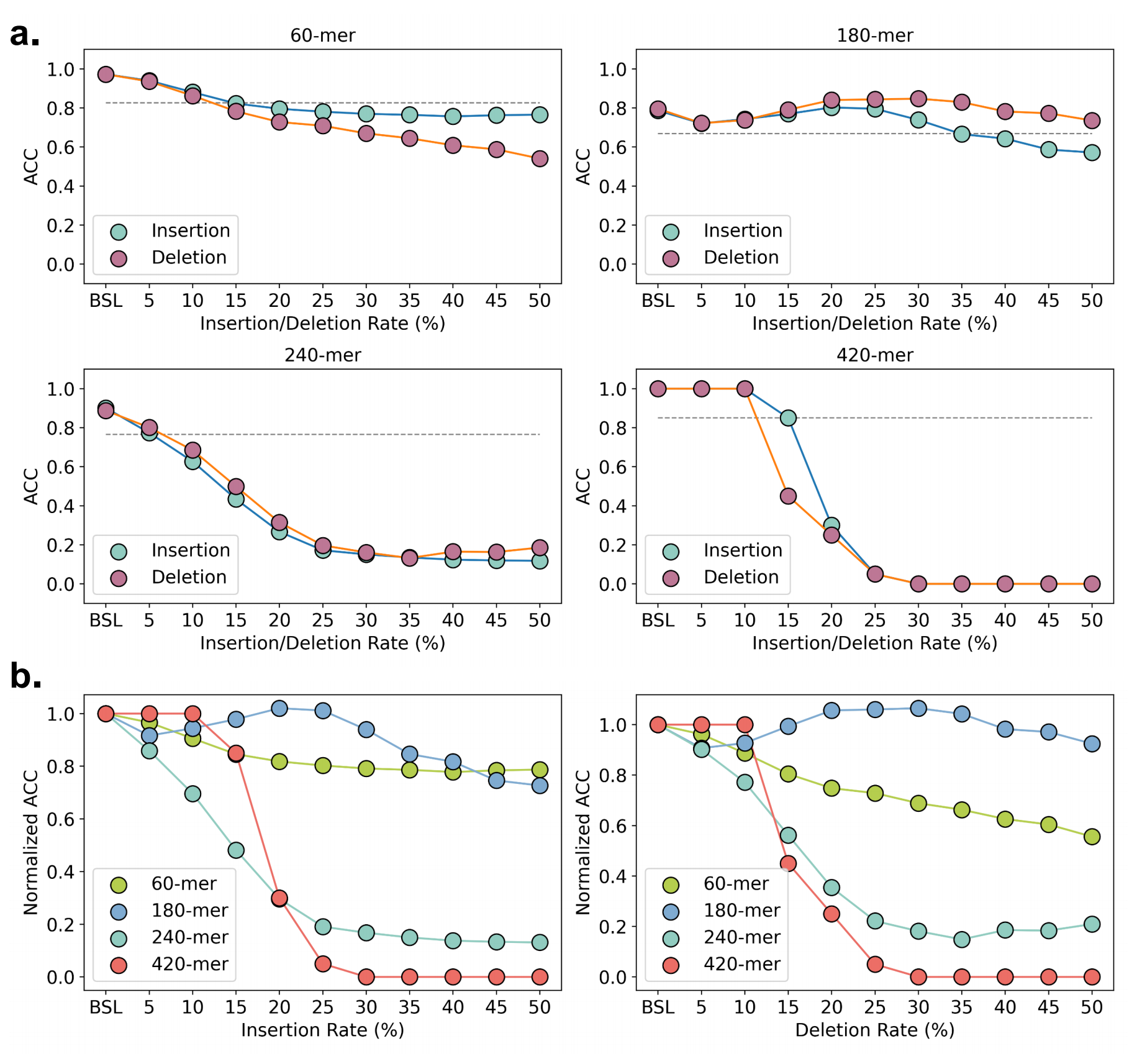}
    \caption{Illustration of VLP structural stability under sequence perturbation. {(a)}: Accuracy curves as a function of insertion and deletion rates across four stoichiometric. {(b)}: Comparison of normalized accuracy curves for the four stoichiometric.   }
    \label{fig:stability}
\end{figure}
 The coat proteins of Q$\beta$ virions may  show high versatility
 and can have diverse forms of  \cite{chang2022structural}.  In the design of VLPs, molecules such as ligands, polysaccharides, and canonical amino acids are often employed to decorate or modify the VLP surfaces to reduce or enhance their immunogenicity or facilitate cell internalization \cite{derdak2006direct,smith2013reengineering}. At the sequence level, such modifications usually require the deletion or insertion of amino acid residues in the original protein sequence.  A key consideration in VLP design is that the modification should preserve the ability of the capsid proteins to self-assemble correctly. Consequently, understanding the structural stability of VLPs under sequence perturbation is important for designing VLPs at the sequence level. Here, we perform the stability analysis under sequence perturbations on the VLP706 dataset based on different stoichiometric. This strategy can also be applied to investigate the structural stability of viruses.

To explore the direct relationship between the VLP sequences and their structural stability, we construct a predictive model by integrating the pretrained protein Transformer ESM-2 \cite{lin2023evolutionary} with the gradient boosting tree algorithm. Specifically, the ESM-2 is used to generate feature embeddings for sequences, and the gradient boosting tree is used for training and prediction. This combined model is referred to as SeqGBT. We train SeqGBT on the full VLP706 dataset and use the resulting model for downstream stability analysis.

Both insertion and deletion operations are considered for simulating sequence perturbation. Specifically, 20 perturbed datasets are generated by randomly inserting or deleting residues in the VLP706 samples at rates of 5\%, 10\%, 15\%, 20\%, 25\%, 30\%, 35\%, 40\%, 45\%, and 50\%. Additionally, two baseline datasets are constructed by inserting or deleting exactly one residue per sequence. We apply the trained SeqGBT model to these newly generated datasets and evaluate the prediction accuracy. To ensure a robust and reliable analysis, the perturbation process is repeated 20 times with different random seeds, and the average ACC of all prediction results is used as the final performance.

The prediction accuracy on these perturbed sequence datasets is shown in Figure~\ref{fig:stability}~(a). As shown in the figure, the accuracy curves for both insertion and deletion operations share similar trends in four stoichiometric. 
Interestingly, the accuracy curve for the 180-mer samples exhibits a non-monotonic trend, increasing and then decreasing with the increase of insertion or deletion rates, distinct from the patterns observed in the other three stoichiometric. Notably, even at a 40\% deletion rate, the 180-mer ACC remains comparable to the baseline, suggesting unique structural robustness that warrants further investigation.
To evaluate structural tolerance, we use the gray dotted line in each subfigure to indicate 85\% of the baseline ACC as a stability threshold. Under this criterion: the 60-mer and 420-mer samples are stable up to 10\% insertion or deletion rate. The 240-mer samples remain stable only up to a 5\% perturbation rate. In contrast, the 180-mer samples maintain stability up tp 35\% insertion or deletion. 

To further compare different stoichiometric, we compute the normalized ACC, defined as the ratio of the ACC on a perturbed dataset to its baseline ACC. The normalized ACC curves are shown in Figure~\ref{fig:stability}~(b). It can be seen that the 180-mer samples exhibit the highest overall robustness to sequence insertion and deletion. The 420-mer samples are stable only under mild perturbation ($\leq$10\%), beyond which accuracy declines sharply. The 240-mer samples are the most sensitive to insertion and deletion, with a rapid ACC loss with the increase of perturbation. The 60-mer samples exhibit good overall stability under insertion rates of up to 50\%, but are slightly more sensitive to deletions.

The findings in this work suggest that 60-mer and 180-mer VLPs are more robust and stable than 240-mer and 420-mer VLPs for the introduction of antibodies and therapeutics for the treatment of viral infection and drug delivery. For the same reason, these VLPs are also more suitable for the delivery of antigens in vaccine development. As an example, bacteriophage Q$\beta$ consists of 180 protein subunits in its icosahedral capsid.  It has been proposed as a robust and multi-purpose carrier to simultaneously deliver multiple SARS‐CoV‐2 variants \cite{tan2024inducing}.  

\section{Discussion}

\begin{figure}[th]
    \centering    \includegraphics[width=0.95\linewidth]{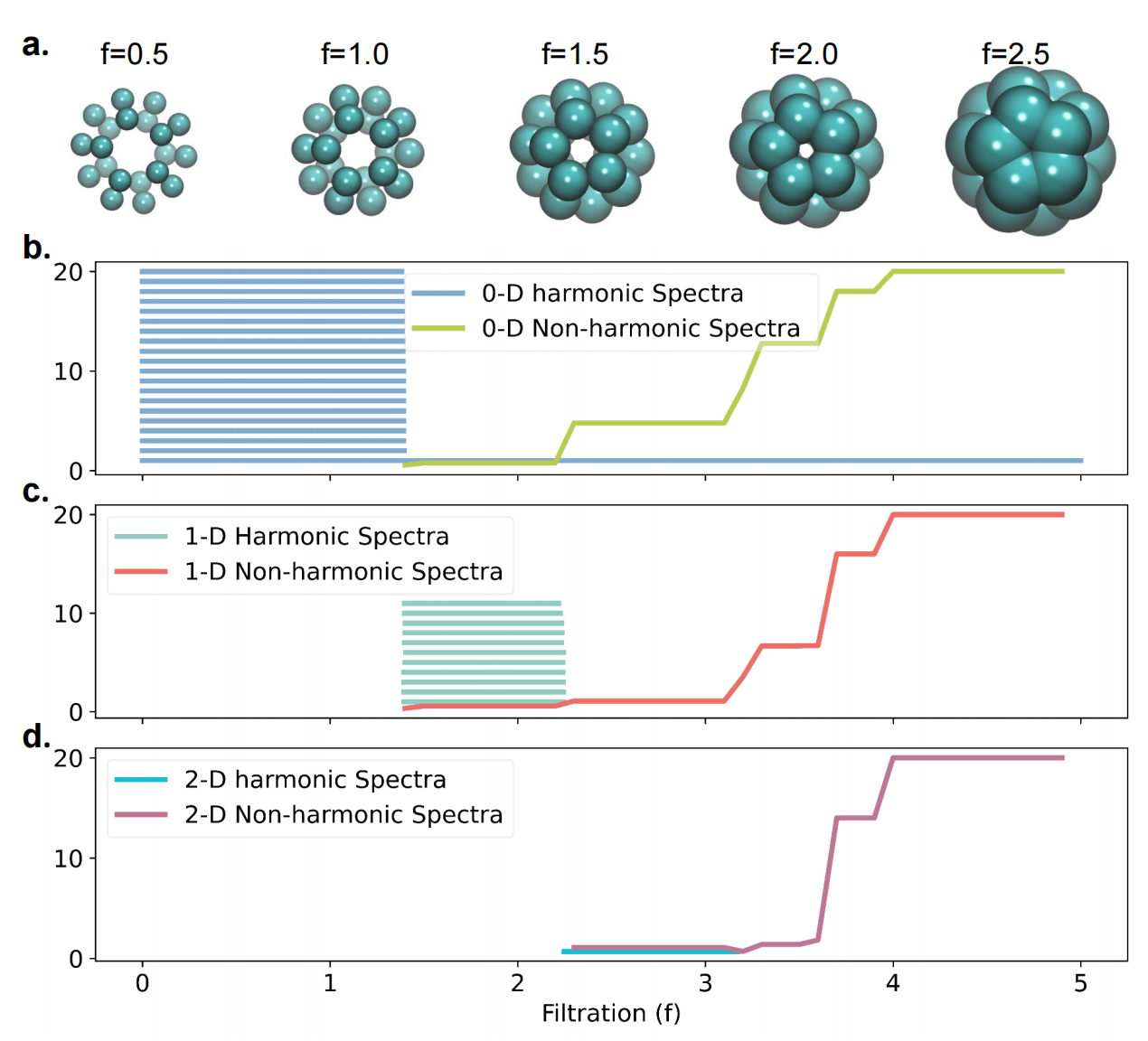}
    \caption{Illustration of persistent Laplacians. (a): a filtration process of the Vietoris-Rips complex constructed from the $C_{20}$. (b): zero-dimensional persistence Laplacian spectra. (c): one-dimensional persistent Laplacian spectra. (d): two-dimensional persistent Laplacian spectra.}
    \label{fig:persistent-laplacian}
\end{figure}
\subsection{Interpretability of Persistent Laplacian}

The spectra of persistent Laplacians encode rich topological and geometric information about the underlying data structures. Here we use the $C_{20}$ as an example to illustrate this interpretability in Figure~\ref{fig:persistent-laplacian}. The $C_{20}$ structure consists of 20 carbon atoms arranged into 12 pentagon patterns each formed by 5 atoms. Figure~\ref{fig:persistent-laplacian}~(a) presents a Vietoris-Rips filtration of the $C_{20}$, while (b), (c), and (d) show the corresponding persistent Laplacian spectra in dimensions 0, 1, and 2, respectively. 

The harmonic spectra of persistent Laplacian at (b), (c), and (d) are isomorphic to persistent homology that captures essential topological features such as connected components, loops, and voids, respectively. In the 0-D harmonic component, the 20 bars correspond to the 20 atoms in the $C_{20}$ structure. In the 1-D harmonic part, the 11 bars reflect the 12 pentagonal loops in the structure (one loop is topologically dependent on the others). The 2-D case contains only 1 bar, corresponding to the global void enclosed by the $C_{20}$ structure. These harmonic spectra efficiently characterize the multiscale topological information of the data.

Beyond topological information, the non-harmonic spectra encode additional geometric information that persistent homology cannot capture. As shown in the figure, the smallest non-zero eigenvalues of persistent Laplacians are plotted for 0-D, 1-D, and 2-D cases in (b), (c), and (d), respectively. In the 0-D and 1-D cases, the eigenvalue curves start at the filtration value of 1.4~\AA, corresponding to the appearance of edges, i.e., the right endpoints of the 19 short bars in (b), which is also the bond length in each pentagon. At this value, the 2-D eigenvalue curve is absent because there is no triangle to form the Laplacian. When the filtration value reaches approximately 2.3 \AA, corresponding to the right endpoints of bars in (c), the 0-D curve begins to rise due to the addition of new edges. At this time, the 1-D curve has a slight increase and the 2-D curve starts, both of which are attributed to the appearance of triangles in the simplicial complex representation. At around 3.2\AA, both the 0-D and 1-D curves exhibit a noticeable jump as numerous edges of similar length are incorporated. Meanwhile, the 2-D curve slightly decreases, as the number of triangles increases more rapidly than that of tetrahedra. As the filtration value increases further, all three curves rise and exhibit jumps at approximately 3.7~\AA and 3.9~\AA, ultimately reaching the value of 20 at 4.0~\AA, which corresponds to the longest pairwise atomic distance in the $C_{20}$ structure. After this point, the three curves remain constant. This is because the simplicial complex has become fully connected: all atoms are connected by edges, all edges are covered by triangles, and all triangles are linked by tetrahedra. The corresponding 0-D Laplacian matrix becomes a square matrix with 19 as diagonal entries and -1 as off-diagonal entries, while both the 1-D and 2-D Laplacian matrices become the diagonal matrices with 20 as diagonal entries, resulting in a uniform non-zero eigenvalue of 20.    

Notably, starting at the filtration value of 3.2~\AA, the persistent homology no longer changes, as there are no further topological changes. However, the addition of higher-dimensional simplices (edges, triangles, tetrahedra) continues to alter simplicial complex induced by the geometric connectivity of the Laplacian model. These geometric features are not captured by persistent homology but are effectively encoded by the persistent Laplacian through its non-harmonic spectral components. 

This example not only offers a detailed geometric and topological explanation of persistent Laplacians but also demonstrates their advantages over persistent homology widely used in TDA. These mathematical models give rise to interpretable machine learning algorithms \cite{papamarkou2024position}.         

\subsection{Reliability of AlphaFold-generated Protein Complex Structures}

AlphaFold, particularly its latest version, offers a relatively robust approach for protein complex structure generation from sequences as accessed in the literature \cite{wee2024evaluation}. However, AlphaFold structures are slightly less accurate than the PDB ones in the prediction of protein-protein binding free energy changes upon mutation. Specifically,  the use of AlphaFold complexes was found to give rise to an  8.6\% increase in the prediction RMSE compared to original PDB complex structures \cite{wee2024evaluation}. Moreover, many AlphaFold complex structures have large errors. These errors were not captured in AlphaFold's ipTM performance metric. In general, AlphaFold complex structures are relatively reliable for crystallizable proteins but are problematic for intrinsically flexible protein regions or domains. As such, we have avoided the use of AlphaFold complex structures as much as we can in our persistent topological Laplacian-based machine learning prediction.    

\section{Methods}
We build a structural subunit-based method for VLP stoichiometry prediction and a sequence-based approach for structural stability analysis. For stoichiometry prediction, we have two datsets: VLP200 and VLP706. The structural subunits (or asymmetric units) for the VLP200 dataset are constructed using the ColabFold implementation \cite{mirdita2022colabfold} of AlphaFold2 \cite{AlphaFold2021} with VLP200 sequence inputs. In contrast, the VLP706 dataset is collected from the RCSB PDB. In this case, AlphaFold2 is applied only to those samples that lack experimentally resolved structures.   
Topological features are then extracted from these structures using the persistent Laplacian and are subsequently used for the machine learning prediction. It is important to note that although AlphaFold2 can predict the 3D structure of a protein from its sequence, it does not provide information about the stoichiometry of VLPs. VLP stoichiometry means the number of protein subunits that form the whole particle. Our structure-based model is specially designed to predict the VLP stoichiometry from its asymmetric subunits.

For the stability analysis under sequence perturbation, we adopt the pretrained protein Transformer ESM2 \cite{lin2023evolutionary} model to directly extract sequence-level features from perturbed samples for use in machine learning tasks. 

In this section, we first give a brief review of the persistent Laplacian for simplicial complexes. We then use $C_{20}$ to illustrate its interpretability. Finally, we give details about the topological feature extraction process applied to VLP structures.
Details about the mathematical theory of persistent Laplacian can be found in Section S1 of the supplementary information, while the machine learning parameters are provided in Section S2.

\subsection{Persistent Laplacian of Simplicial Complexes}

An abstract simplicial complex $K$ over a finite vertex set $V$ is a collection of non-empty subsets of $V$ that is closed under the operation of taking non-empty subsets.
Any element $\sigma\in K$ consisting of $p+1$ vertices is called a $p$-simplex.
Intuitively, a 0-simplex corresponds to a vertex, a 1-simplex to an edge (line segment), a 2-simplex to a filled triangle, a 3-simplex to a tetrahedron, and so on for higher-dimensional counterparts.

An oriented simplicial complex is a simplicial complex equipped with a specific ordering of the vertices. Each simplex inherits an orientation from this global vertex ordering, and a simplex along with its induced vertex order is referred to as an oriented simplex. For simplicity, we will refer to oriented simplices simply as simplices in the remainder of this section.

Let $\mathbb{F}$ be a fixed field coefficient, the $p$-th chain group $C_p(K)$ of $K$ is defined as the vector space over $\mathbb{F}$ whose basis consists of all $p$-simplices in $K$. The boundary operator $\partial_p: C_p(K) \rightarrow C_{p-1}(K)$ is a linear map defined on a $p$-simplex $\sigma = [v_0 v_1 \cdots v_p]$ by
$$\partial(\sigma)=\sum_{i=0}^p(-1)^i[v_0v_1\cdots \hat{v_{i}}\cdots v_p],$$
where $\hat{v_i}$ denotes the omission of the vertex $v_i$.
These boundary operators satisfy the fundamental property $\partial_{p} \circ \partial_{p+1} = 0$, forming a chain complex:
$$\cdots\rightarrow C_{p+1}(K)\xrightarrow{\partial_{p+1}}C_{p}(K)\xrightarrow{\partial_p}C_{p-1}(K)\xrightarrow{\partial_{p-1}}\cdots\xrightarrow{\partial_1}C_0(K)\rightarrow0.$$
The $p$-th homology group is defined as 
$$H_p(K)=Ker\partial_p/Im\partial_{p+1},$$
and its rank, denoted $\beta_p$, is called the $p$-th Betti number. Intuitively, $\beta_0$ represents the number of connected components, $\beta_1$ counts the number of independent loops, and $\beta_2$ measures the number of enclosed voids. 


Given a filtration of a simplicial complex $K$, that is, a nested sequence of its subcomplexes:
$$K_0\subset K_1\subset \cdots \subset K_n=K.$$


Let $C_p^s = C_p(K_s)$. Consider the subspace 
$$C_p^{s,t}=\{\alpha\in C_p^t|\partial(\alpha)\in C_{p-1}^s\},$$
which consists of $p$-chains in $K_t$  whose boundaries lie entirely within $K_s$.
Then we have
$$
C_{p+1}^{s,t}\xrightarrow{\partial^{s,t}_{p+1}}C_p^s\xrightarrow{\partial_p}C_{p-1}^s
$$
where $\partial_{p+1}^{s,t}$ denotes the restriction of the boundary operator $\partial_{p+1}$ to the subspace $C_{p+1}^{s,t}$

The persistent Laplacian from $K_s$ to $K_t$ is defined as
$$L_p^{s,t}=\partial_{p+1}^{s,t}(\partial_{p+1}^{s,t})^*+(\partial_p)^*\partial_p,$$
where $(\cdot)^*$ denotes the adjoint with respect to the standard inner product on chain groups.

A fundamental property of the persistent Laplacian is that its nullity (i.e., the dimension of its kernel) is equal to the persistent Betti number.
Thus, the persistent Laplacian encodes the persistent homology entirely through its harmonic component, while the non-harmonic spectra capture additional geometric and structural information about the evolution of the complex across the filtration.

\subsection{Persistent-Laplacian Vectorization of VLPs}
In our PLML model, each VLP structure is represented as a collection of simplicial complexes, from which persistent Laplacians are computed to extract topological features. Given the large size of typical VLP structures, we restrict our analysis to atoms within the asymmetric unit to reduce computational cost. To construct the simplicial complexes, we adopt the element-specific strategy introduced in \cite{cang2018integration}, in which six atom sets are extracted from each VLP: $\{C_\alpha\}$, $\{N\}$, $\{O\}$, $\{C_\alpha,N\}$, $\{C_\alpha,O\}$, $\{N,O\}$. We construct both Vietoris-Rips complexes \cite{vietoris1927hoheren} and Alpha complexes \cite{edelsbrunner2011alpha} on these atom sets to model their topological structures.

For Vietoris-Rips complexes, we use the standard Euclidean distance and compute the zero-dimensional persistent Laplacian for each of the six atom sets. The harmonic component is derived from a filtration ranging from 0 \AA~ to 10 \AA~ with a step size of 0.5 \AA. A bin-counting method is applied to generate a 20-dimensional feature vector for each atom set. For the non-harmonic components, we use the same filtration and extract seven statistical descriptors from the spectrum of non-zero eigenvalues of the persistent Laplacian: maximum, minimum, mean, sum, standard deviation, variance, and count of eigenvalues. This yields a 140-dimensional feature vector per atom set.

For Alpha complexes, the Euclidean distance is again used for all six atom sets. In this case, we compute the one-dimensional and two-dimensional persistent Laplacians. To simplify computation, only the harmonic component is retained, which corresponds directly to persistent homology. Using barcode representations, we extract the following statistical features: (1) sum, maximum, and mean of bar lengths; (2) maximum and minimum of bar birth times; and (3) maximum and minimum of bar death times. This results in a 14-dimensional feature vector for each atom set. In summary, the total topological feature for a VLP sample is a 1044=(140+20)$\times$6+14$\times$6 dimensional vector. In our implementation, the GUDHI \cite{maria2014gudhi} and Scipy \cite{gommers2024scipy} libraries are utilized to compute persistent Laplacian features.

\section{Conclusions}
Understanding the stoichiometry of VLPs is pivotal for elucidating their assembly efficiency and modulating their immunogenic properties, thereby facilitating vaccine development and optimization. However, existing experimental techniques for determining VLP stoichiometry are often time-consuming and labor-intensive, while current computational approaches remain limited in predictive performance. Consequently, there is a pressing need to develop more efficient and accurate computational methods.
In this work, we present a persistent Laplacian-based topological machine learning model for predicting VLP stoichiometry. By harnessing both the harmonic and non-harmonic spectra of persistent Laplacians, our model captures rich topological and geometric features of VLP structures, offering a comprehensive and robust data representation. This advanced topological method enables our model to outperform existing methods on the VLP200 dataset.
To further evaluate the model’s robustness and generalizability, we construct a larger dataset, VLP706, comprising 706 VLP samples, an expansion in both sample size and stoichiometry types compared to the 200-sample VLP200 dataset. Our model continues to demonstrate strong predictive performance on VLP706, highlighting its effectiveness across diverse VLP structures. We also examined VLP stability by random sequence perturbations. Our study indicates 60-mers and 180-mers are more stable than 240-mers and 420-mers, with a specific order of 180-mers, 60-mers, 420-mers, and 240-mers. This result conforms the advantage of using the bacteriophage Q$\beta$, a 180-mer VLP, for vaccine and drug design \cite{tan2024inducing}.


\section*{Data and Code Availability}
The data and code in this study can be found in  \href{https://github.com/LiuXiangMath/PLML}{github.com/LiuXiangMath/PLML}.

\section*{Acknowledgment}
This work was supported in part by NIH grants R01AI164266, R01AI190348, and R01AI182419, National Science Foundation grant DMS2052983, and Michigan State University Research Foundation.

\section*{Declaration of Interests}
The authors declare no competing interests.


\newpage
\section*{ Supplementary Information  }

\section*{Persistent Laplacian of Simplicial Complexes}

In this section, we provide a brief overview of persistent Laplacians defined on simplicial complexes. We begin by introducing the concepts of simplicial complexes and simplicial homology. We then present the definitions of the combinatorial Laplacian and the persistent Laplacian of simplicial complexes.

\paragraph{Simplicial Complex}
An abstract simplicial complex $K$ over a finite vertex set $V$ is a collection of non-empty subsets of $V$ that is closed under the operation of taking non-empty subsets; that is, for every $\sigma\in K$ and every non-empty subset $\tau\subset \sigma$, it follows that $\tau \in K$. Any element $\sigma\in K$ consisting of $p+1$ vertices is called a $p$-simplex, and its dimension is defined as $p$. The dimension of the complex $K$ is given by the maximum dimension among all its simplices.
From this definition, it is clear that any graph can be viewed as a one-dimensional simplicial complex. Intuitively, a 0-simplex corresponds to a vertex, a 1-simplex to an edge (line segment), a 2-simplex to a filled triangle, a 3-simplex to a tetrahedron, and so on for higher-dimensional counterparts.

An oriented simplicial complex is a simplicial complex equipped with a specific ordering of the vertices. Each simplex inherits an orientation from this global vertex ordering, and a simplex along with its induced vertex order is referred to as an oriented simplex. For simplicity, we will refer to oriented simplices simply as simplices in the remainder of this section.

\paragraph{Simplicial Homology}
Let $\mathbb{F}$ be a fixed field coefficient . For a simplicial complex $K$, the $p$-th chain group $C_p(K)$ is defined as the vector space over $\mathbb{F}$ whose basis consists of all $p$-simplices in $K$. The boundary operator $\partial_p: C_p(K) \rightarrow C_{p-1}(K)$ is a linear map defined on a $p$-simplex $\sigma = [v_0 v_1 \cdots v_p]$ by
$$\partial(\sigma)=\sum_{i=0}^p(-1)^i[v_0v_1\cdots \hat{v_{i}}\cdots v_p],$$
where $\hat{v_i}$ denotes the omission of the vertex $v_i$.
These boundary operators satisfy the fundamental property $\partial_{p} \circ \partial_{p+1} = 0$, forming a chain complex:
$$\cdots\rightarrow C_{p+1}(K)\xrightarrow{\partial_{p+1}}C_{p}(K)\xrightarrow{\partial_p}C_{p-1}(K)\xrightarrow{\partial_{p-1}}\cdots\xrightarrow{\partial_1}C_0(K)\rightarrow0.$$
The $p$-th homology group is defined as 
$$H_p(K)=Ker\partial_p/Im\partial_{p+1},$$
and its rank, denoted $\beta_p$, is called the $p$-th Betti number. Intuitively, $\beta_0$ represents the number of connected components, $\beta_1$ counts the number of independent loops, and $\beta_2$ measures the number of enclosed voids. These Betti numbers serve as fundamental topological invariants that capture the underlying structure of the simplicial complex.

\paragraph{Combinatorial Laplacian of Simplicial Complex}
We equip the chain groups $C_*(K)$ with the standard inner product, defined for any two simplices $\sigma$ and $\tau$ as,
$$<\sigma,\tau>=\delta_{\sigma,\tau},$$
where $\delta_{\sigma,\tau}$ denotes the Kronecker delta.
Let $\partial_p^*$ denote the adjoint of the boundary operator $\partial_p$ with respect to this inner product. The $p$-th combinatorial Laplacian is then defined as
$$L_p=(\partial_p)^*\partial_p+\partial_{p+1}(\partial_{p+1})^*.$$
In matrix representation, $\partial_p^*$ corresponds to the transpose of $\partial_p$. 
A fundamental property of the combinatorial Laplacian is that the nullity of $L_p$  (i.e., the dimension of its kernel) equals the $p$-th Betti number $\beta_p$. Therefore, the Laplacian encapsulates topological information through its harmonic spectra (zero eigenvalues). Moreover, it also encodes rich geometric and combinatorial structure via its non-harmonic spectra, reflecting variations in connectivity and shape beyond homological features.

\paragraph{Persistent Laplacian of Simplicial Complex}
Consider a filtration of a simplicial complex $K$, that is, a nested sequence of its subcomplexes:
$$K_0\subset K_1\subset \cdots \subset K_n=K.$$
Applying the homology functor to this sequence induces a sequence of homology groups connected by homomorphisms:
$$H_p(K_0)\xrightarrow{i_0} H_p(K_1)\xrightarrow{i_1} \cdots \xrightarrow{i_{n-1}} H_p(K_n),$$
where $i_p$ is the homomorphism induced by the inclusion map from $K_p$ to $K_{p+1}$. Let $i_{s,t}$ denote the composition $H_p(K_s)\xrightarrow{i_s}H_p(K_{s+1})\xrightarrow{i_{s+1}}\cdots\xrightarrow{i_{t-1}}H_p(K_t)$, i.e.,
$$i_{s,t}=i_{t-1}\cdots i_s.$$

The $p$-th persistent homology from $K_s$ to $K_t$ is defined as the image
$$H_p^{s,t}=Im(i_{s,t}),$$
representing the homology classes that are born in $K_s$ and are still alive in $K_t$.

To define the persistent Laplacian, let $C_p^s = C_p(K_s)$. Consider the subspace 
$$C_p^{s,t}=\{\alpha\in C_p^t|\partial(\alpha)\in C_{p-1}^s\},$$
which consists of $p$-chains in $K_t$  whose boundaries lie entirely within $K_s$.
Then we have
$$
C_{p+1}^{s,t}\xrightarrow{\partial^{s,t}_{p+1}}C_p^s\xrightarrow{\partial_p}C_{p-1}^s
$$
where $\partial_{p+1}^{s,t}$ denotes the restriction of the boundary operator $\partial_{p+1}$ to the subspace $C_{p+1}^{s,t}$

The persistent Laplacian from $K_s$ to $K_t$ is defined as
$$L_p^{s,t}=\partial_{p+1}^{s,t}(\partial_{p+1}^{s,t})^*+(\partial_p)^*\partial_p,$$
where $(\cdot)^*$ denotes the adjoint with respect to the standard inner product on chain groups.

A fundamental property of the persistent Laplacian is that its nullity (i.e., the dimension of its kernel) is equal to the persistent Betti number:
$${\rm rank}(H_p^{s,t})={\rm rank}(L_p^{s,t})$$
Thus, the persistent Laplacian encodes the persistent homology entirely through its harmonic component, while the non-harmonic spectra capture additional geometric and structural information about the evolution of the complex across the filtration.

\section*{Machine Learning Parameters}

For the machine learning component, the gradient boosting tree algorithm is employed for all datasets. 
   The same hyperparameter configuration is employed for both the structure-based and sequence-based models. The detailed parameters used for this algorithm are provided in Table~\ref{tab:ml}. The scikit-learn package \cite{pedregosa2011scikit} is employed for model training and prediction.
\begin{table}[h]
\centering
\caption{Parameters for gradient boosting tree  }
\label{tab:ml}
\begin{tabular}{|l| c| }
    \hline
    n\_estimators &  4000 \\
    \hline
    max\_depth & 7\\
    \hline
    min\_sample\_split & 3 \\
    \hline
    learning\_rate & 0.01\\
    \hline
    max\_features & sqrt\\
    \hline
    subsample & 0.7\\
    \hline
\end{tabular}
\end{table}

\section*{Evaluation Metrics} 
We provide detailed definitions of all evaluation metrics used in our study. Accuracy is defined as the proportion of correctly classified samples among all samples, and it applies to both binary and multi-class classification tasks. For binary classification, we denote:
\begin{itemize}
    \item $TP$ (True Positives): the number of positive samples correctly predicted as positive
    \item $FP$ (False Positives): the number of negative samples incorrectly predicted as positive
    \item $FN$ (False Negatives): the number of positive samples incorrectly predicted as negative
    \item $TN$ (True Negatives): the number of negative samples correctly predicted as negative
\end{itemize}
Based on these quantities, the following metrics are defined:
$${\rm sensitivity}=\frac{TP}{TP+FN},$$
which measures the proportion of correctly identified positive samples.
$${\rm specificity}=\frac{TN}{TN+FP},$$
which measures the proportion of correctly identified negative samples.
$${\rm precision}=\frac{TP}{TP+FP},$$
which indicates the proportion of correctly predicted positive samples among all predicted positives.
$$NPV=\frac{TN}{TN+FN},$$
which indicates the proportion of correctly predicted negative samples among all predicted negatives.

The AUC \cite{fawcett2006introduction} quantifies the model’s ability to distinguish between positive and negative classes . The ROC curve is generated by plotting the true positive rate against the false positive rate under varying decision thresholds, and the AUC corresponds to the area under this curve.
For multi-class classification, we employ the macro-averaged one-vs-rest AUC. In this approach, each class is considered the positive class while the remaining classes are treated as negative. An AUC score is computed for each class, and the final AUC is obtained by averaging the AUC scores across all classes.


\end{document}